\documentclass[12pt]{article} 
\usepackage{psfig}
\usepackage{a4}

\newcommand{\bra}{\langle}
\newcommand{\ket}{\rangle}
\newcommand{\half}{\frac{1}{2}}

\newcommand{\ep}{\epsilon}

\newcommand{\be}{\begin{equation}}
\newcommand{\ee}{\end{equation}}
\newcommand{\bea}{\begin{eqnarray}}
\newcommand{\eea}{\end{eqnarray}}
\newcommand{\bean}{\begin{eqnarray*}}
\newcommand{\eean}{\end{eqnarray*}}
\newcommand{\nn}{\nonumber}

\begin{document}
\title{
\vskip -100pt
{  
\begin{normalsize}
\mbox{} \hfill \\
\mbox{} \hfill HD-THEP-00-20\\
\mbox{} \hfill March 2000\\
\vskip  70pt
\end{normalsize}
}
 On Thermalization in Classical Scalar Field Theory
\author{
Gert Aarts,\thanks{email: aarts@thphys.uni-heidelberg.de}
\addtocounter{footnote}{1}
Gian Franco Bonini,\thanks{email: bonini@thphys.uni-heidelberg.de}
\addtocounter{footnote}{2} 
and 
Christof Wetterich\thanks{email: C.Wetterich@thphys.uni-heidelberg.de}
\addtocounter{footnote}{3}\\
\normalsize
{\em Institut f\"ur theoretische Physik, Universit\"at Heidelberg}\\
\normalsize
{\em Philosophenweg 16, 69120 Heidelberg, Germany}\\
\normalsize
}
}
\date{March 27, 2000}
\maketitle
 
\renewcommand{\abstractname}{\normalsize Abstract} 
\begin{abstract}
\normalsize 

Thermalization of classical fields is investigated in a $\phi^4$ scalar
field theory in $1+1$ dimensions, discretized on a lattice.  We
numerically integrate the classical equations of motion using initial
conditions sampled from various nonequilibrium probability distributions.
Time-dependent expectation values of observables constructed from the
canonical momentum are compared with thermal ones. It is found that a
closed system, evolving from one initial condition, thermalizes to high
precision in the thermodynamic limit, in a time-averaged sense. For
ensembles consisting of many members with the same energy, we find that
expectation values become stationary - and equal to the thermal values -
in the limit of infinitely many members. Initial ensembles with a nonzero
(noncanonical) spread in the energy density or other conserved quantities
evolve to noncanonical stationary ensembles. In the case of a narrow
spread, asymptotic values of primary observables are only mildly affected.
In contrast, fluctuations and connected correlation functions will differ
substantially from the canonical values. This raises doubts on the use of
a straightforward expansion in terms of 1PI-vertex functions to study
thermalization.

\end{abstract}
 
\newpage

\section{Introduction}

Time evolution of correlation functions in classical field
theory out of thermal equilibrium is of interest for various reasons.
On one side a classical field theory (after discretization on a lattice to
regulate the Rayleigh-Jeans divergence) provides a relatively simple
playground for testing various ideas of statistical mechanics. Examples in
the literature include the approach to thermal equilibrium \cite{Parisi},
the relation with chaos \cite{Heinz}, and the
dynamics in thermal gradients, when the system is coupled to external
heat baths at the boundaries \cite{Aoki}.  A more pragmatic reason is
given by the recent emergence of classical \cite{Khleb,Roos,Juan}
and semiclassical \cite{Aarts} approximations to nonperturbative dynamics
of low-momentum modes in nonequilibrium {\em quantum} field theory.
Knowledge of thermalization properties of classical field theories is
important to determine the applicability of these approaches.

A nonequilibrium ensemble can be defined by an initial probability
distribution with the property that the energy is distributed over the
available degrees of freedom in some nonequilibrium fashion. One would say
that an ensemble with a very large number of degrees of freedom has
``thermalized'' when after sufficiently long time the expectation values
of observables approach the thermal ones, and in fact become
time-independent.  More precisely, we restrict this criterion to
observables and correlation functions with support in a region of space
that is sufficiently small as compared to the total volume. (We will
always consider isolated systems, i.e.\ without an external heat bath.) We
concentrate on equal-time correlation functions, and define thermal values
of correlation functions by expectation values in the canonical ensemble
with the same average energy as the ensemble under consideration.

It has been argued \cite{Wett,fixedpoints,Bonini} that the existence of
conserved correlation functions constitutes an obstruction to
thermalization. This may be illustrated by an ensemble consisting of many
independent experiments with isolated systems where the initial conditions
have some experimental uncertainty in the energy density $\ep$. As an
example, this is realized in heavy-ion collisions by many independent
scattering events. An initial nonvanishing spread $\Delta\ep/\ep$ is
conserved by the time evolution of the ensemble - it corresponds to a
conserved correlation function. In consequence, the ensemble cannot relax
asymptotically to a canonical ensemble with a temperature given by the
average energy density. In the latter case $\Delta\ep/\ep$ would have to
vanish in the thermodynamic limit. In this paper we clarify how this issue
can be related to our understanding of thermalization. 

For our investigation a natural ordering is provided by the type of
ensembles that can be considered. Given an initial nonequilibrium
probability distribution, we call one particular realization of this
distribution a ``microstate''. The simplest ensemble consists of only one
microstate. We will see below that for single microstates fluctuations in
time remain, even for asymptotically large times, and we study therefore
the behaviour of time-averaged observables
\be
 \bra O\ket_{\Delta t}(t) = \frac{1}{\Delta t}\int_{t-\Delta
t}^{t}dt'\, O(t').
\ee
They become stationary for sufficiently large $t$ and $\Delta t$. We shall
call this behaviour ``quasi-stationary''. More advanced ensembles are
formed by superimposing more than one microstate: in particular, an
ensemble consisting of $N_m$ microstates which have the same energy $E$
and contribute with the same weight will be referred to as a fixed-energy
ensemble. Of interest is now the behaviour of ensemble-averaged
observables,
\be
\label{eqOfee}
\bra O\ket_{E,N_m}(t) = \frac{1}{N_m}\sum_{i=1}^{N_m} O_i(t).
\ee
Note that these expectation values can be defined locally in time,
which makes it more suitable for the study of the time evolution of
expectation values out of equilibrium. 
Expectation values in a fixed-energy ensemble with an infinite
number of microstates are denoted as
$\bra O\ket_{E} \equiv \bra O\ket_{E,\infty}$.
Finally, a more general ensemble can be built as the superposition of
states with different energy, weighted with a weight function $f(E)$.

Specific questions can now be raised. Will the nonequilibrium ensembles
described above indeed thermalize according to the definition given
earlier?  Are there restrictions on those ensembles, e.g.\ on the weight
function $f(E)$? Is it important that a fixed-energy ensemble, unlike the
canonical one, has no energy fluctuations? Further: is it possible to
approximate the time evolution using a (truncated) expansion in, for
instance, 1-particle irreducible vertex functions, at arbitrarily long
times?

\section{Classical scalar field theory}

We will address the issues mentioned above from first principles by
numerical simulations using a simple classical scalar field theory in
$1+1$ dimensions with the action
\be
\label{eqaction}
 S = \int dt\,\int_0^L dx \left[ \half (\partial_t\phi)^2 - \half
(\partial_x\phi)^2 - \half m^2\phi^2  - \frac{\lambda}{8}\phi^4\right].
\ee
Here $L$ is the (one-dimensional) volume and we use periodic
boundary conditions. The equation of motion reads $\partial_t^2\phi =
\partial_x^2\phi - m^2\phi -\lambda\phi^3/2$, and the conserved energy is
\be 
E = \int_0^L dx \left[ \half \pi^2 + \half
(\partial_x\phi)^2 + \half m^2\phi^2 +\frac{\lambda}{8}\phi^4\right],
\ee
where $\pi(x,t)=\partial_t\phi(x,t)$ is the canonical momentum. In order
to solve the time evolution, we discretize the action in the standard way
on a lattice with $N$ spatial lattice points, with spatial lattice spacing
$a$ (such that $aN=L$) and time step $a_0$, and derive the equation of
motion from the discretized action.\footnote{This gives the Lagrange
version of the leap frog discretization.} We then solve this equation
numerically, starting from a nonequilibrium initial condition.

Out of the infinite number of possible nonequilibrium initial conditions,
we chose the following microstates: the field $\phi(x,0)$ is initially set 
to zero, and only a few long wavelenghts of $\pi(x,0)$ are excited.
Explicitly, 
\be
\label{eqinit}
 \phi(x, 0) = 0,\;\;\;\;
 \pi(x, 0) = {\sum_{k}}^{\prime} 
A\cos (2\pi kx/L -\psi_k),\label{microstate}
\ee
where $k$ is integer and $0\leq\psi_k<2\pi$ \cite{Parisi}. The prime
indicates that only a very restricted number $n_{\rm e} \ll N$ of modes is
excited. We have chosen $n_{\rm e}=4$ modes with momentum of the order of
the mass ($p \equiv 2\pi k/L\approx m$). For arbitrary coupling constant
the initial energy is $E=n_{\rm e}LA^2/4$. Many initial conditions with a
certain energy can be generated by choosing different phases $\psi_k$ in
Eq.\ (\ref{eqinit}).  We construct a fixed-energy nonequilibrium ensemble
as a superposition of $N_m$ microstates with random $\psi_k$ and equal
weight.  The time evolution of ensemble averages can be calculated by
summing over the realizations. Note that for $N_m\to \infty$, the
fixed-energy ensemble defined in this way is translationally invariant:
for instance the initial two-point function reads
\be 
\bra \pi(x, 0)\pi(y, 0)\ket_E = {\sum_k}^\prime \half A^2\cos
[2\pi k(x-y)/L].\ee
Microstates and ensembles with $N_m<\infty$ do not have this property, of
course.

The class of initial conditions (\ref{eqinit}) is far from classical
thermal equilibrium, where all modes of both $\phi$ and $\pi$ have nonzero
expectation values of their squared amplitude, determined by the
parameters in the action and the temperature $T$. For a free system
($\lambda=0$), the temperature is related to the total energy by
equipartition as $E=NT=LT/a$. Note that this reflects the (linear)
Rayleigh-Jeans divergence in classical thermal field theory, as $a$ goes
to zero.  Corrections due to the nonzero coupling constant are finite in
$1+1$ dimensions.  

It is convenient to use the mass parameter $m$ as the dimensionful scale,
and rescale all dimensionful parameters with $m$, i.e.\ $x=x'/m,\; t=
t'/m,\; \lambda= m^2\lambda'$. If we also rescale the field 
with the dimensionless combination $v\equiv m/\sqrt{3\lambda}$, 
$\phi=v\phi'$, the
classical equation of motion is independent of $\lambda'$.  The
dimensionless energy $E'$ is related to unscaled energy $E$ as 
$E'= E/(mv^2) = 3\lambda E/m^3$, and similar for the temperature:
$E'/T' =E/T$. A given value of $E'$ describes therefore
several values of $\lambda/m^2$ if $E/m$ is changed accordingly. 
Note that for fixed $E/m$ a more strongly coupled theory is
obtained by taking higher $E'$. 
To anticipate the Rayleigh-Jeans divergence in thermal equilibrium, we
keep the ratio $E'/N$ fixed, when varying the lattice spacing or
volume. Besides $E'/N$, the remaining parameters that need to
be specified are the physical volume $L'=mL$, the lattice spacing $a'=ma$
(or equivalently $N=mL/ma$). For the time step we use $a_0/a=0.05$ (we
have also used smaller time steps, and found no significant deviations). 

We concentrate on correlation functions that are built from the conjugate
momentum $\pi(x,t)$, since these can easily be compared with
equal-time
expectation values in thermal equilibrium, where the only nontrivial
correlation function is 
\be \bra \pi(x,t)\pi(y,t)\ket_T = \frac{T}{a}\delta_{xy}.
\ee
With respect to the space
dependence, we consider both expectation values of global observables,
i.e.\ observables averaged over the complete volume, and observables that
are averaged in space over a ``subvolume'' with size $L_s<L$ only. In
the latter case one may think of the rest of the system as an
effective heat bath, though it
is of course still correlated with the subsystem under consideration and
hence not a heat bath in the strict sense.

We focus on expectation values, made from the following two building blocks:
\be G^{(2)}_s(x_s,t) = \frac{1}{N_s}{\sum_x}^{(N_s)}\pi^2(x,t),\;\;\;\;
G^{(4)}_s(x_s,t) = \frac{1}{N_s}{\sum_x}^{(N_s)} \pi^4(x,t).
\ee
The label $N_s$ denotes that the average is taken over a subvolume with
$N_s=L_s/a$ points, centered around $x=x_s$. In case that $N_s=N$ the
complete volume average is taken, and the subscript $s$ is omitted.
The kinetic energy in a subvolume is given by
\be
K_s(x_s,t)=\half aN_s G^{(2)}_s(x_s,t).
\ee 
In thermal equilibrium its value is $\bra K_s\ket_T=N_sT/2$, independent
of $x_s$ of course. One may define an effective temperature, locally in
time and out of equilibrium, from the expectation value
\begin{equation}
\label{eqtemp}
T_{\pi} (t)= 2\bra K\ket(t)/N.
\end{equation}
Note that this is an example of a ``primary'' quantity, i.e.\ it is
directly given as an ensemble average. 

The observables we investigate are the following.
In the canonical ensemble $\pi$ is a gaussian (free) field, which implies 
that $\bra G^{(4)}_s\ket_T = 3\bra G^{(2)}_s\ket_T^2$. We want to see
whether $\pi$ becomes eventually gaussian in the nonequilibrium ensembles
as well. We define the deviation from $\pi$ being gaussian in a
subvolume as
\be 
\mbox{dev}(\pi_s) = \frac{ \bra G_s^{(4)}(x_s, t) \ket}{3\bra
G_s^{(2)}(x_s, t)\ket^2}-1.
\label{eqdevpi}
\ee
For one microstate or in a fixed-energy ensemble the conserved total energy 
has no fluctuations. The time evolution of the fluctuations in the kinetic 
energy of subsystems 
\be
 \mbox{fluc}(K_s) = \frac{ \bra (K_s - \bra K_s\ket)^2\ket}{\bra
K_s\ket^2}.
\ee 
is interesting for two reasons: first, although they
never  vanish, for a closed system one expects some suppression
from the fact that the total energy is conserved. Second,
one can calculate in the canonical case the size of the fluctuations 
exactly: $\mbox{fluc}(K_s)_T = 2/N_s$. This provides the possibility for a 
quantitative comparison.
The last observable we discuss is the normalized third moment of the
kinetic energy in a subvolume (the second moment is related to
fluc($K_s$)), and is denoted as
\be 
\mbox{mom}(K_s^3) =  \frac{\bra K_s^3\ket}{\bra K_s\ket^3}.
\ee
In thermal equilibrium mom$(K_s^3)_T=1+6/N_s+8/N_s^2$.  This observable
differs from the previous ones in that it is neither directly related to
fluctuations nor purely local, since it involves the expectation value of
$\bra \pi^2(x,t)\pi^2(y,t)\pi^2(z,t) \ket$ with $x,y,z$ all in the same
subvolume. These three observables are related in a nonlinear
manner to ensemble averages, and hence examples of ``secondary'' quantities.
Note also that in the canonical ensemble they are independent of the  
temperature.

\begin{figure}
\centerline{\psfig{figure=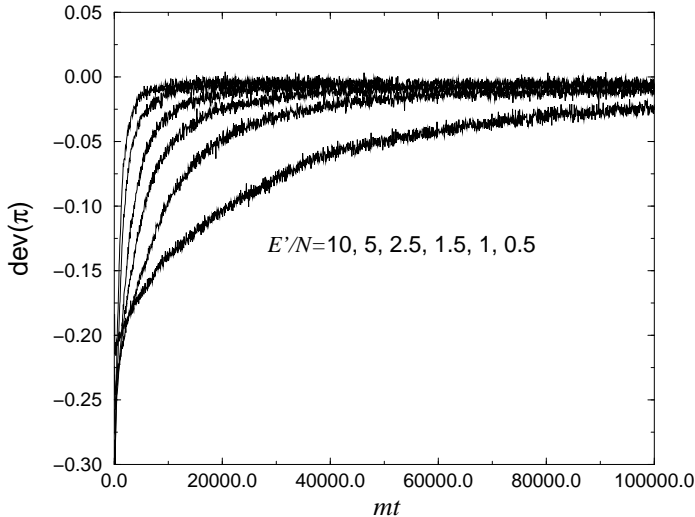,height=7cm}}
\caption{Time dependence of dev($\pi)_{\Delta t}$, averaged over $m\Delta 
t = 62.5$, for different values of the energy $E'/N$, 
from $E'/N=10$ (top curve, fastest relaxation) to $E'/N=0.5$
(bottom curve, slowest relaxation).
Each curve represents the average over 20 independent initial conditions
to reduce fluctuations. Parameters are $mL=32, N=256$. 
}
\label{figapproach1}
\end{figure}       

\section{Single microstates}

We start with ensembles consisting of one microstate only. 
For each $E'/N$ we generate a number of initial conditions with the
same energy, and integrate the equation of motion. We take time
averages over an interval $\Delta t$ for each microstate separately. 
For convenience we sometimes average at the end over independent runs 
in order to reduce the fluctuations in time.

\begin{figure}
\centerline{\psfig{figure=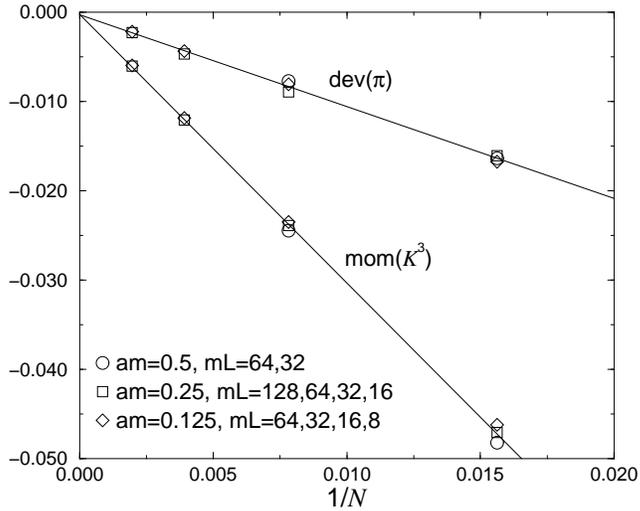,height=7cm}}
\caption{Dependence of dev($\pi)_{\Delta t}$ and the relative deviation of
mom($K^3)_{\Delta t}$ from the thermal value on the number of
lattice points $N$, in the quasi-stationary regime with $m\Delta
t=10000$, for various values of $mL$ and $ma$. 
No significant dependence on the subvolume size or $E'/N$ has been
found. The lines are straight-line fits and statistical errors are smaller
than the symbol sizes. 
}
\label{figdevpivsN}
\end{figure}      

To get some feeling for the time scales involved, we show in Fig.\
\ref{figapproach1} how dev($\pi)_{\Delta t}$ relaxes towards zero as a
function of time, for various values of the energy. In this case the time
averages are taken over an interval $m\Delta t=62.5$, and since the size
of the fluctuations in time in each microstate remains rather large, we
show in Fig.\ \ref{figapproach1} curves averaged over 20 independent
initial conditions per energy value. 
As stated before, there is a correspondence between high energy and large
coupling, and we see that the time scale involved increases
for decreasing $E'/N$, as expected. 

It is clear from Fig.\ \ref{figapproach1} that $\pi$ tends to become
gaussian: dev($\pi)_{\Delta t}\to 0$. To make this precise, we discard
the initial part where  dev($\pi)_{\Delta t}$ is not quasi-stationary 
and  calculate the time average in a long,
interval $m\Delta t=10000$, for several values of $mL$
and $ma$, subvolume sizes $mL_s=4, 8, 16$, and energies 
$E'/N = 2,\ldots, 10$. For each $E'/N$ we have generated 10 initial
conditions. We have observed no significant dependence of the long-time values
on either the initial condition for a fixed energy, the energy $E'/N$
in the interval indicated, or the subvolume size. The asymptotic average
value of  dev($\pi)_{\Delta t}$ depends only on the total number of degrees of
freedom $N$,
as shown in Fig.\ \ref{figdevpivsN}.
Note that changing $mL$ with fixed $am$ or vice versa has the same effect.
For finite $N$,  we see that $\pi$ deviates from being gaussian,
on the order of a percent for $N=100$. 
However, in the thermodynamic limit, which is $N\to\infty$ in this
classical lattice model, $\pi$ becomes a gaussian field.  A
straight-line fit shows that the
thermodynamic limit is approached as dev($\pi)_{\Delta t}=-1.03(2)/N-0.0003(2)$, where the
numbers between brackets indicate the error in the last digit from the  
fit. 
Also shown in Fig.\ \ref{figdevpivsN} is the relative deviation of 
mom($K^3)_{\Delta t}$ from the thermal value. 
Again we found no significant dependence on the subvolume size, the energy,
or the particular initial condition. Also in this case we see a 
deviation from the thermal value for finite $N$, which vanishes in the
thermodynamic limit as $-3.01(3)/N-0.0003(3)$.

\begin{figure}
\centerline{\psfig{figure=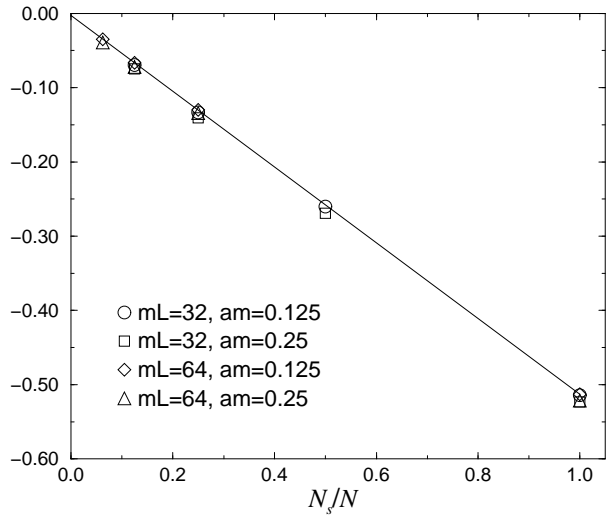,height=7cm}}
\caption{Relative deviation of fluc($K_s)_{\Delta t}$ from the thermal
value, as a function of $L_s/L=N_s/N$, with $mL_s=4,8,16$, for
various values of $mL$ and $ma$. The line is a straight-line fit through
the $mL=64, am=0.125$ points. 
}
\label{figK2vsN}
\end{figure}       

The relative deviation of $\mbox{fluc}(K)_{\Delta t}$ from the thermal
value is shown in Fig.\ \ref{figK2vsN}, again in the quasi-stationary
regime with $m\Delta t=10000$, using several
energies as before. We present the result as a function of $L_s/L=N_s/N$.
For $N_s/N=1$, we see that the fluctuations are suppressed by
approximately
50\%, compared to the thermal value. A straight-line fit shows that the
data are consistent with 
\be
\mbox{fluc}(K_s)_{\Delta t}=\frac{2}{N_s}\left(1-\frac{1.4(2)}{N} -
0.514(1)\frac{N_s}{N}\right).
\ee
In other words, in the proper thermodynamic limit, $N\to\infty$ and 
$N_s/N\to 0$, the fluctuations are given by the canonical ones.
On the other hand, in the limit $N\to\infty$ with
{\em fixed} $N_s/N$, $\mbox{fluc}(K_s)_{\Delta t}$  deviates from 
the canonical value, confirming the expectation that the heat bath 
must be infinitely larger than the subsystem under consideration.

The results obtained so far  indicate that single microstates evolving from
an initial condition taken from the class (\ref{eqinit}) thermalize in
the thermodynamic limit, in a time-averaged sense. 

\section{Fixed-energy ensembles}

\begin{figure}
\centerline{\psfig{figure=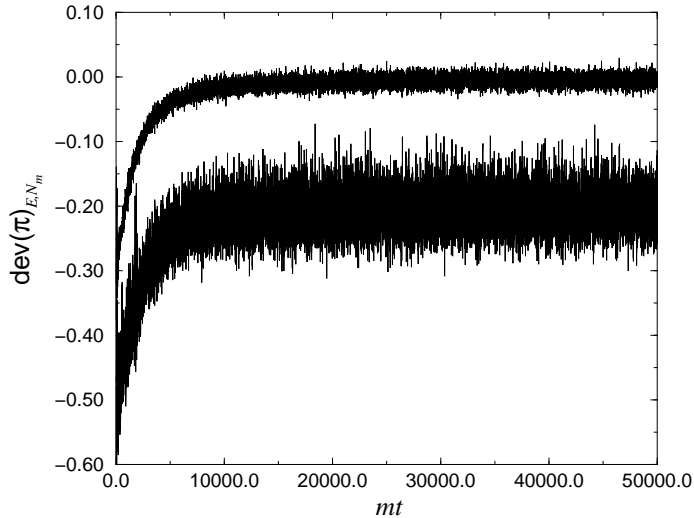,height=7cm}}
\caption{Time dependence of dev($\pi)_{E,N_m}$ in nonequilibrium ensembles
with $N_m=10$ members (lower curve, shifted down by 0.2 for clarity) and
$N_m=150$ members (upper curve). In a larger ensemble the size 
of fluctuations is reduced. Parameters are $mL=32, N=256, E'/N=4$. 
}
\label{figfluc_10_150}
\end{figure}       

\begin{figure}
\centerline{\psfig{figure=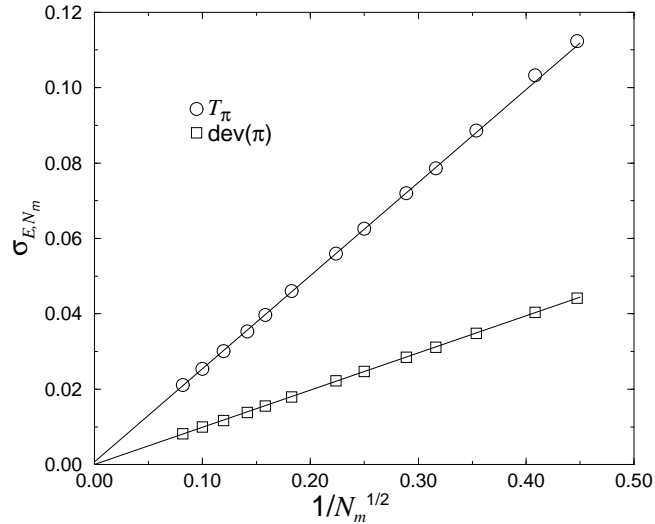,height=7cm}}
\caption{Size of the fluctuations $\sigma_{E,N_m}$ for $T_\pi$ and
dev($\pi)$ versus 
$1/\sqrt{N_m}$, 
where $N_m$ denotes the number of
members in a fixed-energy ensemble, ranging from 5 to 150. 
The lines are straight-line fits through the eight lowest points. 
Same parameters as in Fig.\ \ref{figfluc_10_150}.}
\label{figfluc_Nin}
\end{figure}

To extend the analysis, we continue with fixed-energy ensembles, built
as  superpositions of $N_m>1$ microstates (\ref{microstate})
 with the same energy $E$. We
present the effect of ensemble averaging for dev($\pi)_{E,N_m}$ in Fig.\
\ref{figfluc_10_150} for ensembles consisting of $N_m=10$ and $N_m=150$
members, respectively. We emphasize that there is no time averaging 
performed here, unlike
the microstate case. It is seen that  the larger the ensemble, the 
smaller the size of
the fluctuations. To make this quantitative, we identify the
size of the fluctuations with the standard deviation
$\sigma_{E,N_m}$ around the mean value in the  quasi-stationary regime.
Explicitly, for a generic primary observable 
\be
\label{eqvar}
\sigma^2_{E,N_m} \equiv 
\Big\bra \bra O\ket_{E,N_m}^2\Big\ket_{\Delta t} -
\Big\bra \bra O\ket_{E,N_m}\Big\ket_{\Delta t}^2.
\ee
In Fig.\ \ref{figfluc_Nin} we show the asymptotic values of $\sigma_{E,N_m}$ 
for the observables $T_\pi$ and 
dev($\pi)$, as functions of $1/\sqrt{N_m}$. For the
quasi-stationary regime the interval $30000<mt<50000$ was used.
We see that in an ensemble with $N_m\to\infty$ members
the standard deviations vanish, which implies that the expectation
values become time-independent, and the ensemble becomes stationary.

%The  vertical axis is intercepted at $0.0007\pm 0.0002$, and $0.00002\pm
%0.00001$ respectively. 

These results are consistent with the results
obtained in ensembles consisting of one microstate only.  Using  definition
(\ref{eqOfee}) for a primary observable $O$, we find that the
variance (\ref{eqvar}) is given by
\bea
\nn
\sigma^2_{E,N_m} &=&
\frac{1}{N_m^2}\sum_{i,j}\left[ \bra O_iO_j\ket_{\Delta t}(t) -
\bra O_i\ket_{\Delta t}(t)\bra O_j\ket_{\Delta t}(t)\right]\\
\nn
&=&
\frac{1}{N_m^2}\sum_{i}\left[ \bra O_i^2\ket_{\Delta t}(t) -
\bra O_i\ket^2_{\Delta t}(t)\right]\\
&=& \frac{1}{N_m^2}\sum_{i}\sigma^2_{\Delta t, i},
\eea
where $\sigma_{\Delta t, i}$ denotes the standard deviation around the mean
value in the quasi-stationary regime in the $i$th microstate.  
The second line follows if  the individual microstates are statistically
independent, i.e.\ $\bra O_iO_j\ket_{\Delta t}(t)=\bra O_i\ket_{\Delta 
t}(t) \bra O_j\ket_{\Delta t}(t)$ for $i\neq j$. If we now use that the
standard
deviation $\sigma_{\Delta t, i}$ in a particular realization is independent
of the specific details of the initial condition, such that
$\sigma_{\Delta t, i} \equiv \sigma_{\Delta
t}$ for all $i$, we conclude that $\sigma_{E,N_m} = \sigma_{\Delta
t}/\sqrt{N_m}$, in agreement with the numerical results.
Note that this implies that the initial probability distribution given by
(\ref{eqinit}) with $n_{\rm e}=4$ is already ``rich'' enough to define a
fixed-energy ensemble with stationary asymptotic behaviour.

\section{Generic ensembles}

Perhaps surprisingly, we can use the results obtained above  to 
show that generic  ensembles approach noncanonical stationary distributions.  
Let us consider an initial ensemble whose  energy is not precisely
fixed, but has some finite spread: $\Delta E^2 \equiv \bra (E-\bra
E\ket)^2\ket\neq 0$. Note that the spread is conserved during time evolution.
The issue of thermalization in this case is more complicated. To get
some feeling for this, let us 
first consider the simple case of a superposition of 
fixed-energy ensembles weighted with a normalized gaussian function $f(E)$:
\begin{equation}
\label{eqfE}
f(E) = {\cal N}_{\kappa}\exp\left[-\half\kappa(E-\bar E)^2/\bar
E^2\right],
\end{equation}
with ${\cal N}_\kappa=(\kappa/2\pi\bar E^2)^{1/2}$ and where $\kappa$ is a
large but finite constant in the thermodynamic limit. The energy spread
for this ensemble is $\Delta E^2/\bar E^2 =1/\kappa$.  Under the
assumption that the individual fixed-energy ensembles evolve to the
corresponding microcanonical ones, the large-time asymptotic stationary
values of all primary quantities can be obtained by folding the
microcanonical ones with $f(E)$. It is then straightforward to compute
also secondary quantities such as connected functions. Using definition
(\ref{eqtemp}) for the temperature $T_\pi$, we 
obtain\footnote{Neglecting the nonlinear $T$-dependence of $E$.}
\bea
\label{devpi_gaussian}
&&\mbox{dev}(\pi)_\kappa\approx 1/\kappa,\\
&&
\bra\pi^2(x)\pi^2(y)\ket_\kappa -
\bra\pi^2(x)\ket_\kappa\bra\pi^2(y)\ket_\kappa
\approx (T_\pi/a)^2/\kappa,
\;\;\;x\neq y.
\label{cluster}
\eea
We emphasize that secondary quantities defined for the whole ensemble are {\em not}
weighted averages of the corresponding quantities in  the individual
microcanical ensembles.
{}From Eq.\ (\ref{devpi_gaussian}) we see that this ensemble does not have
thermal correlation functions, i.e.\ dev($\pi)_{\kappa}\neq 0$, not even in the
thermodynamic limit. Furthermore, since the left-hand-side of Eq.\
(\ref{cluster}) is nothing else than the connected four-point function and
the right-hand-side does not vanish for $|x-y|\to\infty$, we find that the
ensemble (\ref{eqfE}) does not obey the clustering property \cite{Weinberg}.
A violation of clustering has a rather large impact on other observables,
for instance for $N\gg N_s$ we find that
\be
\mbox{fluc}(K_s)_\kappa =
\frac{2}{N_s}\left(1+\frac{N_s+2}{2\kappa}\right).
\ee
Fluctuations in the kinetic energy in subvolumes will deviate
substantially from the canonical value when $\kappa \sim N_s$. In the
limit that $N_s\gg \kappa$, the size of the fluctuations approaches
$1/\kappa$ and therefore does not vanish.

In order to make the discussion more general, let us now consider a system 
with $n_c$ conserved {\em intensive}
quantities  ${\bf c}=\{c_\mu\}$, $\mu=1,\ldots,n_c$. Supported by our
numerical results,  we assume 
that all microstates  with given ${\bf c}$  evolve, in 
an asymptotic, time-averaged sense, to the same final state, i.e.\ that
all
details of the initial state other than  the conserved quantities are lost.
In other words, for a local observable $O$ in microstate $i$:
\begin{equation}
\lim_{t\rightarrow\infty,\Delta t\rightarrow \infty,\Delta t/t\rightarrow 0}
\frac{1}{\Delta t}\int_{t-\Delta t}^t O_i(t)=
\langle O\rangle^*_{N,{\bf  c}}.
\end{equation}
Here $\langle O\rangle^*_{N,{\bf  c}}$ is the expectation value of $O$ in
the microcanonical ensemble with fixed ${\bf c}$ and $N$ 
degrees of freedom.\footnote{Asymptotic values will generically be
denoted with a star.} 
We further assume that an ensemble consisting of  a superposition
of sufficiently many states, all with the same values of 
${\bf c}$,
evolves to the corresponding microcanonical ensemble:
\begin{equation}
\lim_{t\rightarrow\infty}\langle O(t)\rangle_{\bf c}= 
\langle O\rangle^*_{N,{\bf c}}. \label{eq:fixed_e_ensemble}
\end{equation}
Notice that in this case no time averaging is required.

We then consider ensembles that are superpositions of states with different 
values of ${\bf c}$, weighted with a normalized weight function $f_N({\bf
c})$. As an example, one may consider again a gaussian distribution:
\[
f_N({\bf c})=\prod_{\mu}\frac{1}{\sqrt{2\pi\Delta c_{\mu}^2}}
\exp\left[ -\half(c_{\mu}-\overline{c}_{\mu})^2/\Delta c_{\mu}^2\right],
\]
with $\Delta c_{\mu}\sim 1/N^{\alpha}$. 
In this case $\Delta {\bf c}^2 \equiv
\overline{{\bf c}^2}-{\overline{\bf c}}^2\neq 0$, where $\overline{\bf c}=
\int d{\bf c}\,f_N({\bf c}){\bf c}$.
Provided that the ensemble is sufficiently 
``rich'', i.e.\ all subensembles with given values of 
${\bf c}$ satisfy Eq.\ (\ref{eq:fixed_e_ensemble}), the  
asymptotic expectation values depend only on the weight function
$f_N({\bf c})$:
\begin{equation}
\langle O\rangle^*_{f_N}=\int d{\bf c}\;
f_N({\bf c})\langle O\rangle^*_{N,{\bf c}}.
\label{eq:generic_ensembles}
\end{equation}
For $\Delta c_\mu \sim 1/N^\alpha$ we extract the leading
$N$-dependence by writing the weight function $f_N$ as
\begin{equation}
f_N({\bf c})=N^{\alpha n_c}\varphi({\bf s}),\;\;\;\;
{\bf s}=N^{\alpha}({\bf c}-{\bf \overline{c}}), 
\label{eq:N-scaling}
\end{equation} 
where we assume that $\varphi$ is in first approximation
$N$-independent.
The case $\alpha=0$ corresponds to fixed $\Delta c_{\mu}/c_{\mu}$, 
independent of $N$. For a canonical ensemble 
one has $f^{\rm can}_N(E/N)=Z_T^{-1}\Omega(E)e^{-E/T}$, where 
$\Omega(E)$ is the number of states with energy density 
$\epsilon=E/N$ and $Z_T$ is the canonical partition function. In this case,
the spread in the energy density is given by $\Delta\epsilon\sim N^{-1/2}$
and therefore $\alpha=1/2$. In general, we shall assume $\alpha\geq 0$.

We expand $\langle O\rangle^*_{N,{\bf c}}$ in a Taylor series around 
${\bf\overline{c}}$ such that 
Eq.\ (\ref{eq:generic_ensembles}) yields
\begin{eqnarray}
\langle O\rangle^*_{f_N}&=&
\sum_{m=0}^{\infty}\frac{1}{m!N^{\alpha m}}\frac{\partial^m
\langle O\rangle^*_{N,{\bf\bar c}}}{\partial c_{\mu_1}\ldots\partial
c_{\mu_m}}
\int d{\bf s}\,\varphi({\bf s})s_{\mu_1}\ldots s_{\mu_m}\nonumber\\
&=&\langle O\rangle^*_{N,{\bf\bar c}}+
\frac{\overline{s_{\mu}s_{\nu}}}{2N^{2\alpha}}\frac{\partial^2
\langle O\rangle^*_{N,{\bf\bar{c}}}}{\partial c_{\mu}\partial c_{\nu}}
+{\cal O}(N^{-3\alpha}),
\label{eq:N-exp}
\end{eqnarray}
with $\overline{s_{\mu}s_{\nu}} = 
\int d{\bf s}\,\varphi({\bf s})s_{\mu}s_{\nu}$.
Notice that the $m=1$ term in the series drops out due to the definition of 
${\bf\overline{c}}$.
In the case that $f_N$ is the weight function for the canonical ensemble, 
Eq.\ (\ref{eq:N-exp}) shows that the expectation values in
a fixed-energy ensemble will differ from the canonical ones as $\sim
1/N$. This is exactly the behaviour we found numerically (see above).
Generic ensembles with $\alpha>0$ can be compared with the canonical one
by applying (\ref{eq:N-exp}) twice, i.e.\ by subtracting from
(\ref{eq:N-exp}) the same equation with $f_N=f^{\rm can}_N$.
It is then found that primary observables, i.e.\ expectation values that 
are directly given as ensemble averages, approach the canonical
values in the thermodynamic limit, and the way this limit is approached is
directly proportional to the spread in the conserved quantities
$\overline{s_{\mu}s_{\nu}}$.
For a finite $\Delta c_{\mu}/c_{\mu}$ in the thermodynamic limit, the
asymptotic expectation values will differ by a nonzero amount from both
the microcanonical and the canonical ensembles, even in the thermodynamic
limit. However, this difference is suppressed by $\Delta c^2$.

\begin{figure}
\centerline{\psfig{figure=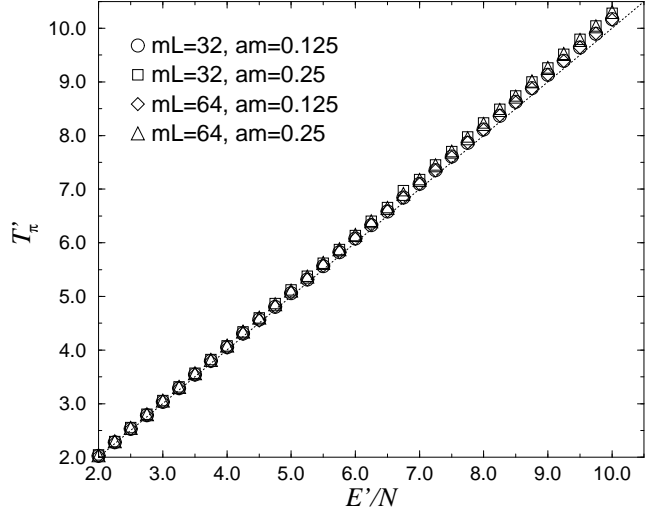,height=7cm}}
\caption{
Dependence of the effective temperature $T'_{\pi}$ on the energy density
$E'/N$, for several  values of $ma$ and $mL$. The dashed line is the
free ($\lambda=0$) relation.
}
\label{fig:temp}
\end{figure}

For secondary quantities, i.e.\ observables that are 
nonlinear functions of ensemble averages, the situation is more involved.
For definiteness, let us go back to the scalar theory 
(\ref{eqaction}), which has only one conserved quantity: 
$c_1=\epsilon$.\footnote{The other obvious candidate, the total momentum
$P = \sum_x\pi(x) \partial_x\phi(x)$, is not conserved on the lattice. We 
have checked that expectation values of $P$ and $P^2$
approximately thermalize, regardless of their initial values.}
We restrict ourselves to translation-invariant ensembles. Notice also that 
the connected two-point functions for $\phi$ and $\pi$ behave 
asymptotically as primary quantities, since 
$\langle \phi \rangle^*_{N}=\langle \pi \rangle^*_{N}=0$.
An example of a secondary quantity is the equal-time connected 4-point
correlation function
\begin{eqnarray}
\langle \psi(x)\psi(y)\psi(z)\psi(w)\rangle_{f_N,C}
&\equiv& \langle
\psi(x)\psi(y)\psi(z)\psi(w)\rangle_{f_N}\label{eq:connected}
\\
&&-\left(\langle \psi(x)\psi(y)\rangle_{f_N}\langle
\psi(z)\psi(w)\rangle_{f_N}
      + \mbox{\em two perm.}\right),
\nonumber
\end{eqnarray} 
with $\psi=\{\pi,\phi\}$.
Here we have extended the standard definition of connected correlation
functions to arbitrary ensembles.
In particular, $\bra\pi^4(x)\ket^*_{f_N,C}$ is related to
dev($\pi)$ in Eq.\ (\ref{eqdevpi}). Using Eq.\ (\ref{eq:N-exp}), we find
\be
\label{eqdevpi2}
\bra \pi^4(x)\ket^*_{f_N,C} \approx  \bra \pi^4(x)\ket^*_{N,\bar{\ep},C} 
+ 
\frac{\overline{s^2}}{2N^{2\alpha}}
\left[ 
\frac{d^2}{d\ep^2}\bra\pi^4(x)\ket^*_{N,\bar{\ep},C}  
+ 6
\left(\frac{d}{d\ep}\bra\pi^2(x)\ket^*_{N,\bar{\ep}}\right)^2\right],
\ee
with $\overline{s^2}/N^{2\alpha}=\Delta \epsilon^2$.
In the case that $f_N=f_N^{\rm can}$ (with $\alpha=1/2$), the
connected 4-point function on the left-hand-side vanishes,
and the fluctuations in the energy density are
given by $\overline{s^2}=T^2\partial\bar\ep/\partial T$. If we use the
definition of temperature in the fixed-energy ensemble
$T_\pi/a=\bra\pi^2(x)\ket^*_{N,\bar{\ep}}$ and
work consistently to order $1/N$, so that $T_\pi$ can be identified with 
the temperature $T$ in the canonical ensemble, Eq.\ (\ref{eqdevpi2})
yields 
\be
\bra \pi^4(x)\ket^*_{N,\bar{\ep},C} =
-3\frac{\overline{s^2}}{N}
\left(\frac{d}{d\ep}\bra\pi^2(x)\ket^*_{N,\bar{\ep}}\right)^2
= -\frac{3}{N}\left(\frac{T}{a}\right)^2\frac{\partial T}{\partial\bar\ep},
\ee
and 
\be  \mbox{dev}(\pi)_{E} = 
\frac{\bra\pi^4(x)\ket^*_{N,\bar{\ep},C}}{3\bra\pi^2(x)\ket^{*2}_{N,\bar{\ep}}}
= -\frac{1}{N}\frac{\partial T}{\partial\bar\ep}. \label{eq_dtde}
\ee
We have measured dev$(\pi)$ and the energy dependence of the
effective temperature in fixed-energy ensembles independently 
(Figs.\ \ref{figdevpivsN} and \ref{fig:temp}, respectively), and have found 
Eq.\ (\ref{eq_dtde}) to be satisfied within numerical accuracy.

As we saw in the example (\ref{eqfE}),  the connected function  
(\ref{eq:connected}) does not, in general, obey the clustering property.  
Using Eq.\ (\ref{eq:N-exp}), we can write
\begin{eqnarray}
\langle \psi^2(x)\psi^2(y)\rangle^*_{f_N,C} 
&\approx&
\langle \psi^2(x)\psi^2(y)\rangle^*_{N,\bar\epsilon,C}
+
\frac{\overline{s^2}}{2N^{2\alpha}} 
\left[ \frac{d^2}{d\epsilon^2}
\langle \psi^2(x)\psi^2(y)\rangle^*_{N,\bar\epsilon, C}\right.
\nonumber\\
\label{eqpsipsi}
&&
\left.
+2\left( 
\frac{d}{d\epsilon}\langle\psi^2(x)\rangle^*_{N,\bar\epsilon}\right)^2
+4\left(
\frac{d}{d\epsilon}\langle\psi(x)\psi(y)\rangle^*_{N,\bar\epsilon}\right)^2
\right]. 
\end{eqnarray}
The limit of large spatial separation for the fixed-energy ensembles can be extracted
by comparison with the canonical ensemble, where clustering holds and the 
left-hand-side vanishes for $|x-y|\to \infty$. Using $f_N=f_N^{\rm can}$,
one finds
\be
\lim_{|x-y|\rightarrow\infty} 
\langle \psi^2(x)\psi^2(y)\rangle_{N,\bar\epsilon,C}^* =
-\frac{T^2}{N}\frac{\partial\overline\epsilon}{\partial T}
\left(\frac{d}{d\epsilon}\langle\psi^2(x)\rangle^*_{N,\bar\epsilon}\right)^2
+ {\cal O}(N^{-3/2}).
\label{eq:4-point}
\ee
In the microcanonical ensemble the connected 4-point  function
vanishes, in the limit of large separations, only up to terms of order
$1/N$. 
Using this result in Eq.\ (\ref{eqpsipsi}) for an generic ensemble with
a nonzero (noncanonical) spread in the energy density, we find that 
also in these ensembles the clustering property will be
violated. For $\alpha=0$ the right-hand-side of Eq.\ (\ref{eqpsipsi}) is
dominated by the first term on the second line and we recover the result
(\ref{cluster}).

It is instructive to translate these findings to momentum space.
We introduce the notation
\begin{equation}
\langle\psi(q_1)\ldots\psi(q_n)\rangle\equiv
Na\delta_{q_1+\ldots+q_N,0}\langle\langle\psi_1\ldots\psi_n\rangle\rangle
\end{equation}  
where we factored out a Kronecker delta function times the volume to take
translational invariance into account. For a clustering system,
all reduced connected functions $\langle\langle...\rangle\rangle$ 
should be finite in the limit $N\to\infty$, $a$ fixed.
For the connected 4-point function of a generic ensemble we have
\begin{eqnarray}
\langle\langle\psi_1\psi_2\psi_3\psi_4\rangle\rangle_{f_N,C}^*
\!\!\!&=&\!\!\!
\langle\langle\psi_1\psi_2\psi_3\psi_4\rangle\rangle_{f_N}^*
-
Na\left(\langle\langle\psi_1\psi_2\rangle\rangle_{f_N}^*
\langle\langle\psi_3\psi_4\rangle\rangle_{f_N}^*\delta_{q_1,-q_2} +
\mbox{\em perm.}
\right)
\nonumber\\
\!\!\!&=&\!\!\!
\langle\langle\psi_1\psi_2\psi_3\psi_4\rangle\rangle_{N,\bar\epsilon,C}^*
+\frac{\overline{s^2}}{2N^{2\alpha}}\frac{d^2}{d\ep^2}
\langle\langle\psi_1\psi_2\psi_3\psi_4\rangle\rangle_{N,\bar\ep,C}^*
\nonumber\\
&&\!\!\!
\!\!\!
+ \frac{\overline{s^2}a}{N^{2\alpha-1}}
\left[
\frac{d}{d\epsilon}
\langle\langle\psi_1\psi_2\rangle\rangle_{N,\bar\epsilon}^*
\frac{d}{d\epsilon}
\langle\langle\psi_3\psi_4\rangle\rangle_{N,\bar\epsilon}^*\delta_{q_1,-q_2}
+ \mbox{\em perm.}\right].
\end{eqnarray}
When momenta come in pairs, the reduced connected 4-point
functions for the microcanonical and canonical ensembles 
differ even in the thermodynamic limit, and, remarkably, for
ensembles with $\alpha<1/2$ these functions actually diverge. 
A similar calculation for the connected 6-point function gives, to leading
order,
\bea 
\bra\bra \psi_1\ldots\psi_6\ket\ket^*_{f_N,C} &=& \bra\bra
\psi_1\ldots\psi_6\ket\ket^*_{N,\bar{\ep},C} \\
\nonumber
&&\hspace{-2.5cm}+
\frac{\overline{s^2}a^2}{N^{2\alpha-2}}
\left[
\bra\bra\psi_1\psi_2\ket\ket^*_{N,\bar\ep} 
\frac{d}{d\ep}\bra\bra\psi_3\psi_4\ket\ket^*_{N,\bar\ep}
\frac{d}{d\ep}\bra\bra\psi_5\psi_6\ket\ket^*_{N,\bar\ep}
\delta_{q_1,-q_2}\delta_{q_3,-q_4}
+ \mbox{\em perm.}\right].
\eea
A comparison with the canonical ensemble shows that the reduced connected
6-point function in the microcanonical ensemble diverges in the
thermodynamic limit $\sim N$. For higher order connected functions to be
finite for all momentum combinations, $f_N$ has to agree with $f^{\rm
can}_N$ to higher and higher order in $1/N$. Approximation methods based
on expansions in connected or 1-particle irreducible correlation functions
may break down since higher-point functions could strongly affect the
expansion. Without taking care of the failure of clustering properties
they may not be suitable to describe thermalization phenomena. This might
explain some qualitatively different results obtained in
\cite{fixedpoints,Bonini}, where the same scalar theory was studied using
truncations of an exact flow equation for the time evolution of equal-time
correlation functions \cite{Wett}. It also puts doubts on the validity of
approximations which lead to thermalization for arbitrary initial
ensembles. 

\section{Conclusion}

Let us summarize our conclusions for the time evolution of classical
ensembles with nonequilibrium initial conditions, based on numerical
results for the $\phi^4$ theory in $1+1$ dimensions on the lattice. The
investigated statistical ensembles approach in the large-time limit
stationary ensembles that depend only on the probability distribution for
the energy density (and the other intensive conserved quantities). For
ensembles that consist of a small number of microstates, the asymptotic
ensemble is approached only in a time-averaged sense.  For a very large
number of degrees of freedom (thermodynamic limit) we find that ensembles
with a nonzero energy density spread do not approach the canonical
ensemble: their large-time primary observables differ from the thermal
values by a finite amount proportional to the spread. Ensembles with
vanishing spread, on the other hand, do thermalize, up to terms suppressed
by the number of degrees of freedom.  These corrections can be computed,
and our numerical results were found to be in excellent agreement with the
analytical results.

We also notice that generic statistical ensembles do not cluster for large
time. Secondary quantities (like connected or 1PI functions) receive
corrections that in coordinate space do not vanish in the limit of large
spatial separation; as a result, corrections to quantities defined over
large subvolumes can be very large. In momentum space, violation of
clustering implies that for specific momentum values secondary quantities
diverge in the thermodynamic limit.  Clustering requires a vanishing
initial spread in all conserved quantities with extreme precision.  We
argue therefore that care should be exercised when applying approximate
methods based on expansions in connected or 1-particle irreducible
correlation functions to the study of thermalization: such methods might
in fact break down in the large-time regime.

\vspace{5mm}

{\bf Acknowledgements}

\noindent{This work was supported by the TMR network {\em Finite Temperature Phase
Transitions in Particle Physics}, EU contract no.\ FMRX-CT97-0122.}

\end{document}